\documentstyle[prb,aps]{revtex}
\draft
\begin{document}
\title{Exactly solvable toy models of unconventional magnetic alloys:\\
Bethe Ansatz versus Renormalization Group method}
\author{Valery I. Rupasov\cite{em}}
\address{Department of Physics, University of Toronto, Toronto,
Ontario, Canada M5S 1A7\\
and\\
Landau Institute for Theoretical Physics, Moscow, Russia}
\date{\today}
\maketitle
\begin{abstract}
We propose toy models of unconventional magnetic alloys,
in which the density of band states, $\rho(\epsilon)$,
and hybridization, $t(\epsilon)$, are energy dependent;
it is assumed, however, that $t^2(\epsilon)\propto\rho^{-1}(\epsilon)$,
and hence an effective electron-impurity coupling
$\Gamma(\epsilon)=\rho(\epsilon)t^2(\epsilon)$ is energy
independent. In the renormalization group approach, the physics
of the system is assumed to be governed by $\Gamma(\epsilon)$
only rather than by separate forms of $\rho(\epsilon)$ and
$t(\epsilon)$. However, an exact Bethe Ansatz solution of
the toy Anderson model demonstrates a crucial role of a form
of inverse band dispersion $k(\epsilon)$.
\end{abstract}


The Kondo problem in ``unconventional'' Fermi systems, where
an effective density of states of band electrons vanishes
either precisely at the Fermi level (``gapless'' systems) or
on some interval around the Fermi level (``gapped'' systems),
has been attracting a significant theoretical interest. Using
poor-man's scaling, Withoff and Fradkin \cite{WF} have found
that the Kondo effect in gapless systems takes place only if
an effective electron-impurity coupling exceeds some critical
value. Numerical renormalization group (RG) calculations,
large-$N$ studies, and quantum Monte Carlo simulations of the
gapless \cite{BH,CF,CJ,I,GBI,BPH} and gapped
\cite{S,SO,OS,TSS,TSGS,CPC,CJ2} systems have confirmed this
prediction and revealed a number of new additional features of
the physics of unconventional magnetic alloys.

In a conventional metallic system with (i) a linear dispersion
of band electrons near the Fermi level and (ii) an energy
independent electron-impurity coupling, basic impurity models
are exactly solved by the Bethe ansatz (BA) \cite{TW,AFL,PS,AH}.
It is well known that the Wilson's numerical RG \cite{W,KM} and
BA methods lead to identical results. Moreover, renormalizability
of the Kondo and Anderson models proved respectively
by Abrikosov \cite{A} and Haldane \cite{H} has been reproven in
the course of a BA solution. Therefore, it is reasonable to
study the Kondo problem in unconventional Fermi systems, where
the BA technique cannot be straightforwardly used, by making
use of scaling arguments and the numerical RG approach.

However, it has recently been found \cite{R} that integrability
of the $U\to\infty$ nondegenerate and degenerate Anderson models
is not destroyed because of a nonlinear dispersion of particles
and an energy dependent hybridization, but it becomes only hidden
\cite{RS}. The developed BA approach has allowed us to study
\cite{R2} the thermodynamics of a $U\to\infty$ Anderson impurity
in a BCS superconductor, and can be used to obtain an exact
solution of the Kondo problem in other unconventional Fermi
systems.

The results obtained \cite{R,R2} demonstrate a discrepancy
between the RG and BA approaches to the Kondo problem in
unconventional Fermi systems. In the RG approach, the
separate forms of the density of states, $\rho(\epsilon)$,
and an energy dependent hybridization, $t(\epsilon)$, are
assumed to be {\em unimportant}. It is assumed thus that
the impurity properties are governed only by an effective
particle-impurity coupling
$\Gamma(\epsilon)=\rho(\epsilon)t^2(\epsilon)$. However,
BA equations contain both an effective coupling and an
inverse band dispersion $k(\epsilon)$. The latter describes
the spatial behavior of wave functions and enters BA
equations via periodic boundary conditions imposed on
system's eigenfunctions. A form of inverse dispersion
plays an extremely important role in the physics of the
system.

In this paper, we study a toy model describing an Anderson
impurity embedded in a gapless Fermi system. In the toy model,
the density of states and hybridization are energy dependent.
It is assumed, however, that $t^2(\epsilon)=2\Gamma\rho^{-1}(\epsilon)$,
where $\Gamma=\mbox{const}$, and an effective coupling is thus
energy independent. Therefore, in the framework of RG approach,
physical properties of the toy and Anderson models are obviously
identical.

We assume a simple form of the density of states,
\begin{equation}
\rho(\epsilon)=\frac{|\epsilon|^r}{|\epsilon|^r+\beta^r},
\end{equation}
where the energy $\epsilon$ is taken relative to the Fermi
level. The parameter $\beta$ determines the size of a region
with an unconventional behavior of the density of states.
At $\beta=0$, we come back to the Anderson model. In the
region near the Fermi level, $|\epsilon|\ll\beta$, the
density of the states exhibits a power-law variation,
$\rho(\epsilon)\sim|\epsilon|^r$.

The toy model is obviously integrable. It is clear also that
a corresponding integrable toy version can be constructed in
the described-above manner for any conventional integrable
impurity model. The physical properties of the toy versions
of the exchange models, such as the s-d (Kondo) and
Cogblin-Schrieffer models, are not affected by a form of
dispersion, because in these models charge and spin excitations
of a system are decoupled from each other \cite{TW,AFL,PS,AH}.
The physics of the toy versions of the Anderson models, where
charge and spin excitations strongly interact, is shown to be
governed by a form of inverse dispersion $k(\epsilon)$.

We start with the Hamiltonian of a nondegenerate Anderson impurity
written in terms of the Fermi operators $c^\dagger_\sigma(\epsilon)$
($c_\sigma(\epsilon)$) which create (annihilate) an electron with
a spin $\sigma=\uparrow,\downarrow$ in an $s$-wave state of energy
$\epsilon$,
\begin{mathletters}
\begin{equation}
H=H_c+H_d+H_h\\
\end{equation}
where
\begin{eqnarray}
H_c&=&\int_{-D}^{D}\frac{d\epsilon}{2\pi}\epsilon
c^\dagger_\sigma(\epsilon)c_\sigma(\epsilon)\\
H_d&=&\epsilon_d d^\dagger_\sigma d_\sigma +
Ud^\dagger_\uparrow d_\uparrow d^\dagger_\downarrow d_\downarrow\\
H_h&=&\int_{-D}^{D}d\epsilon\sqrt{\Gamma(\epsilon)}
[c^\dagger_\sigma(\epsilon)d_\sigma+d^\dagger_\sigma c_\sigma(\epsilon)].
\end{eqnarray}
are the conduction band, impurity and hybridization terms,
respectively. All notation in Eqs. (2) are standard. The
electron energies and momenta are taken relative to the Fermi
values, which are set to be equal to zero. The integration
over the energy variable $\epsilon$ is restricted by the
band half width $D$. In what follows, we assume that $D$
is the largest parameter on an energy scale, $D\to\infty$.
In the energy representation, the effective particle-impurity
coupling $\Gamma(\epsilon)=\rho(\epsilon)t^2(\epsilon)$ combines
the density of band states, $\rho(\epsilon)=(d\epsilon(k)/dk)^{-1}$,
and the energy dependent hybridization $t(\epsilon)$.

Here we consider a toy version of the model assuming that
$t^2(\epsilon)=2\Gamma\rho^{-1}(\epsilon)$,
where $\Gamma=\mbox{const}$. The toy model is obviously
integrable at an arbitrary $U$. Indeed, let us introduce
the Fourier images of the electron operators,
$$
c_\sigma(\tau)=\int\frac{d\epsilon}{2\pi}
c_\sigma(\epsilon)\exp{(i\epsilon\tau)},
$$
and transform thus Eqs. (2) to the auxiliary $\tau$ space.
In this space related to the particle energy, the model
Hamiltonian coincides with the Hamiltonian of the standard
Anderson model written in terms of the operators in the
auxiliary $x$ space,
$$
c_\sigma(x)=\int\frac{dk}{2\pi} c_\sigma(k)\exp{(ikx)},
$$
related to the particle momentum. Therefore, in the $\tau$
space the $N$-particle wave functions of the system,
$\Phi_{\sigma_1\ldots\sigma_N}(\tau_1,\ldots,\tau_N)$,
are given by the standard Bethe ansatz formulae derived by
Wiegmann \cite{PW}. However, periodic boundary conditions
must be imposed on a wave function
$\Psi_{\sigma_1\ldots\sigma_N}(x_1,\ldots,x_N)$ on an interval
of size $L$ in the $x$ space rather than in the $\tau$ space.
These different representations of wave functions are related by
\end{mathletters}
\begin{mathletters}
\begin{equation}
\Psi_{\sigma_1\ldots\sigma_N}(x_1,\ldots,x_N)=
\int_{-\infty}^{\infty}d\tau_1\ldots\tau_N\prod_{j=1}^{N}u(x_j|\tau_j)
\Phi_{\sigma_1\ldots\sigma_N}(\tau_1,\ldots,\tau_N),
\end{equation}
where the ``dressing'' function $u(x|\tau)$ is found to be
\begin{equation}
u(x|\tau)=\int_{-\infty}^{\infty}\frac{dk}{2\pi}
\left(\frac{d\epsilon(k)}{dk}\right)^{1/2}\exp{[i(kx-\epsilon(k)\tau)]}.
\end{equation}
In the Anderson model, where $\epsilon(k)=k$, the dressing function
is nothing but the Dirac delta function, $u(x|\tau)=\delta(x-\tau)$,
and hence the $x$ and $\tau$ representations coincide.

In what follows, we confine ourselves to the case of $U\to\infty$.
Then, imposing periodic boundary conditions on the wave function
$\Psi_{\sigma_1\ldots\sigma_N}(x_1,\ldots,x_N)$ results in the
following BA equations:
\end{mathletters}
\begin{mathletters}
\begin{eqnarray}
e^{ik_jL}\frac{\omega_j-\epsilon_d/2\Gamma-i/2}
{\omega_j-\epsilon_d/2\Gamma+i/2}&=&\prod_{\alpha=1}^{M}
\frac{\omega_j-\lambda_\alpha-i/2}
{\omega_j-\lambda_\alpha+i/2}\\
\prod_{j=1}^{N}\frac{\lambda_\alpha-\omega_j-i/2}
{\lambda_\alpha-\omega_j+i/2}
&=&-\prod_{\beta=1}^{M}
\frac{\lambda_\alpha-\lambda_\beta-i}{\lambda_\alpha-\lambda_\beta+i}
\end{eqnarray}
where $M$ is the number of particles with spin ``down'',
$k_j\equiv k(\omega_j)$, and $\omega=\epsilon/2\Gamma$.
The particle-impurity and effective particle-particle
scattering amplitudes coincide with those in the Anderson
model, because they are really determined only by the
effective particle-impurity coupling $\Gamma$, as it is
assumed in the RG approach. However, the BA equations
contain also the phase factors $\exp{(ik_jL)}$ accounting
for the spatial behavior of wave functions. A form of the
inverse dispersion $k(\omega)$ is clear to play a very
important role in both further mathematical BA constructions
and in the physics of the system.

As in the Anderson model, in the thermodynamic limit spin
``rapidities'' $\lambda_\alpha$ are grouped into bound spin
complexes of size $n$,
\end{mathletters}
\begin{equation}
\lambda_\alpha^{(n,j)}=\lambda_\alpha+i(n+1-2j)/2,\;\;\;
j=1,\ldots,n
\end{equation}
Apart from charge excitations with real charge ``rapidities''
$\omega_j$, the spectrum of the system contains also charge
complexes with complex rapidities
\begin{mathletters}
\begin{equation}
\omega_\alpha^{(\pm)}=\lambda_\alpha\pm i/2,
\end{equation}
provided the signs of the imaginary parts of $\omega_\alpha^{(\pm)}$
and corresponding momenta $k^{(\pm)}=k(\omega_\alpha^{(\pm)})$ are
the same,
\begin{equation}
\mbox{sign}(\mbox{Im}\,k^{(\pm)})=\mbox{sign}(\mbox{Im}\,\omega^{(\pm)}).
\end{equation}
In the Anderson model, due to the linear dispersion law
$k=2\Gamma\omega$ the necessary condition (NC) (6b) is
obviously satisfied at an arbitrary $\lambda\in(-\infty,\infty)$.
In the toy model, a solution of NC is governed by a form
of $k(\omega)$.

To solve Eq. (6b), one has to specify first the power $r$
in the expression for the density of states (1). Here, we
consider two most important cases $r=1$ and $r=2$. At $r=2$
the inverse dispersion is found to be
\end{mathletters}
\begin{mathletters}
\begin{equation}
\frac{k}{2\Gamma}=\omega-\delta\arctan{\frac{\omega}{\delta}},
\end{equation}
where $\delta=\beta/2\Gamma$. Solving Eq. (6b), we find
a critical value of the parameter $\delta$,
\begin{equation}
\exp{(1/\delta_{cr})}=\frac{\delta_{cr}+1/2}{\delta_{cr}-1/2}.
\end{equation}
At $\delta<\delta_{cr}$, NC has a solution at all $\lambda$.
At $\delta>\delta_{cr}$, NC has no solution on the interval
$G_\Delta=(-\Delta,\Delta)$, where
\begin{equation}
\Delta^2=\frac{2\delta}{1-\exp{(-2/\delta)}}-(\delta+1/2)^2.
\end{equation}
Thus, in a sharp contrast to the Anderson model, the spectrum
of charge complexes of the system at $\delta>\delta_{cr}$
contains a gap of size $2\Delta$. The gap appears at
$\delta=\delta_{cr}$, and grows with increasing $\delta$.
At very large $\delta$, $\delta\gg 1$, the gap asymptotically
reaches the maximal value $2\Delta_{max}=1/\sqrt{3}$.

In a gapless Fermi system with the power $r=1$, the inverse
dispersion of band states is found to be
\end{mathletters}
\begin{equation}
\frac{k}{2\Gamma}=\omega+\delta\ln{\frac{\delta-\omega}{\delta}},
\;\;\; \omega<0.
\end{equation}
Solving Eq. (6b), we find no gap in the spectrum of
charge complexes. Thus, the toy model with the power
$r=1$ is qualitatively equivalent to the Anderson model.
Therefore, in what follows we confine ourselves to the toy
model with $r=2$.

The thermodynamics of the system is described by a set of basic
equations for the renormalized energies of elementary excitation
$\varepsilon(\omega)$, $\xi(\lambda)$ and $\kappa_n(\lambda)$,
corresponding to unpaired charge excitations with real $\omega$,
charge complexes and spin complexes of size $n$, respectively:
\begin{mathletters}
\begin{eqnarray}
\varepsilon(\omega)&=&2\Gamma\omega+
a_1*F[-\xi(\omega)]-\sum_{n=1}^{\infty}a_n*F[-\kappa_n(\omega)]\\
\xi(\lambda)&=&4\Gamma\lambda
+a_1*F[-\varepsilon(\lambda]+a_2*F[-\xi(\lambda)]\\
\kappa_n(\lambda)&=&\sum_{m=1}^{\infty}A_{nm}*F[-\kappa_m(\lambda)]
+a_n*F[-\varepsilon(\lambda)].
\end{eqnarray}
In these equations, $F[f(x)]\equiv T\ln{[1+\exp{(f(x)/T)}]}$,
$a_n(x)=(2n/\pi)(n^2+4x^2)^{-1}$, and $A_{nm}(x)=a_{|n-m|}(x)
+a_{n+m}(x)+2\sum_{k=1}^{{\rm min}(n,m)-1}a_{|n-m|+2k}(x)$.
The symbol $*$ stands for the convolution of functions, e. g.
\end{mathletters}
\begin{equation}
a_n*F[\xi(\lambda)]\equiv\int d\lambda' a_n(\lambda-\lambda')
F[\xi(\lambda')].
\end{equation}
For the $U\to\infty$ Anderson model, the thermodynamic
BA equations were derived by Schlottmann \cite{PS2}.
They also can be obtained by setting $U\to\infty$
in the thermodynamic BA equations of the general
Anderson model \cite{TW}. In the toy model, the only
difference is the appearance of the gap in the spectrum
of charge complexes at $\delta>\delta_{cr}$. Therefore,
the integration contour $C$ in the integrals with the
function $\xi(\lambda)$ consists of two intervals,
$C=(-\infty,-\Delta)\oplus(\Delta,\infty)$.

The physical properties of the ground state of the system
are governed by the gap size. At $\Delta=0$, the ground
state, as in the Anderson model, is composed of charge
complexes only. They occupy all states from $\lambda=-D/2\Gamma$
to $\lambda=Q$, where $Q$ is determined by the condition
$\xi(Q)=0$. In conventional metallic systems $Q$ is a large
negative value \cite{TW}, $Q=-(1/2\pi)\ln{(D/\Gamma)}$.
The ground state of the toy model is not affected by the
gap until the gap size does not exceed $|Q|$. When $\Delta$
exceeds $\Delta_{cr}=|Q|$, the ground state of the system is
reconstructed: charge complexes with $\lambda\in(-\Delta,Q)$
decay into unbounded charge excitations and spin waves.

Therefore, in the continuous limit the BA equations (4)
describing the ground state of the system take the form
of integral equation for the ``particle'', $\rho(\omega)$
($\rho(\omega)=0$ at $\omega>B$), $\sigma(\lambda)$
($\sigma(\lambda)=0$ at $\lambda>\Delta$), $\eta(\lambda)$,
and ``hole'', $\tilde{\rho}(\omega)$ ($\tilde{\rho}(\omega)=0$
at $\omega<B$), $\tilde{\sigma}(\lambda)$ ($\tilde{\sigma}(\lambda)=0$
at $\lambda<-\Delta$), densities of distributions of charge
excitations, charge complexes and spin waves, respectively,
\begin{mathletters}
\begin{eqnarray}
\frac{1}{2\pi}\frac{dk(\omega)}{d\omega}+
\frac{1}{L}a_1(\omega-\frac{\epsilon_d}{2\Gamma})&=&
\rho(\omega)+\tilde{\rho}(\omega)+a_1*\sigma(\omega)
+a_1*\eta(\omega)\\
\frac{1}{2\pi}\frac{dq(\lambda)}{d\lambda}+
\frac{1}{L}a_2(\lambda-\frac{\epsilon_d}{2\Gamma})&=&
\sigma(\lambda)+\tilde{\sigma}(\lambda)+a_1*\rho(\lambda)
+a_2*\sigma(\lambda)\\
a_1*\rho(\lambda)&=&\eta(\lambda)+a_2*\eta(\lambda),
\end{eqnarray}
where $q(\lambda)=k(\lambda+i/2)+k(\lambda-i/2)$ is the
momentum of the charge complexes. The ``Fermi level'' of
unbounded charge excitations, $B$, is found from the condition
\end{mathletters}
\begin{equation}
\frac{N}{L}=\int_{-\infty}^{B}d\omega\rho(\omega)+
2\int_{-\infty}^{-\Delta}d\lambda\sigma(\lambda).
\end{equation}
In Eqs. (11), the densities can be separated into the host and
impurity parts, e. g. $\rho(\omega)=\rho_h(\omega)+L^{-1}\rho_i(\omega)$.
The population of the impurity level, $n_d$, and the impurity
spin $S^z_i$ are then given by
\begin{mathletters}
\begin{eqnarray}
n_d&=&\int_{-\infty}^{B}d\omega\rho_i(\omega)+
2\int_{-\infty}^{-\Delta}d\lambda\sigma_i(\lambda)\\
S^z_i&=&\frac{1}{2}\int_{-\infty}^{B}d\omega\rho_i(\omega)-
\int_{-\infty}^{\infty}d\lambda\eta_i(\lambda),
\end{eqnarray}

It is easy to see that Eqs. (11) and (13) are analogous to
equations describing the ground state of the Anderson model
in an external magnetic field ${\cal H}$. The latter, however,
contain no spin waves, $\eta(\lambda)=0$. Therefore,
the impurity spin is given by
\end{mathletters}
$$
S^z_i=\frac{1}{2}\int_{-\infty}^{B'}d\omega\rho _i(\omega),
$$
where the limit $B'$ is determined by the field ${\cal H}$.
In the limit ${\cal H}\to 0$, $B'\to -\infty$, and the impurity
spin vanishes.

In the toy model, unbounded charge excitations and spin waves appear
in the ground state of the system in the absence of an external
magnetic field due to a decay of charge complexes. However, it
is easy to show from Eqs. (11c) and (13b) that their contributions
to the impurity spin precisely compensate each other. Thus, the
Kondo effect takes place in the toy $U\to\infty$ Anderson model
with the powers $r=1$ and $r=2$ at any effective coupling $\Gamma$
and gap size.

At $\Delta<|Q|$, unbounded charge excitations and spin waves
disappear from the ground state, and the population of the impurity
level is determined by a single function $\sigma_i(\lambda)$,
$n_d=2\int_{-\infty}^{Q}d\lambda\sigma_i(\lambda)$. As in the
Anderson model, the impurity population is governed by
the renormalized impurity level energy
$\epsilon^*_d=\epsilon_d+2\Gamma Q$. At $\Delta>|Q|$,
the limit $Q$ is replaced by $-\Delta$, and moreover
a contribution of unpaired charge excitations appears
in Eq. (13a). Therefore, we should expect essential
changes in the behavior of the impurity population and
the impurity magnetic susceptibility compared to the
Anderson model.

In summary, we presented an exact BA analysis of the toy
version of a model describing a $U\to\infty$ Anderson
impurity embedded in a gapless Fermi system. In the RG
approach, the physics of the system is assumed to be
governed by an effective electron-impurity coupling only.
Therefore, the toy and Anderson models should be identical
to each other. The BA analysis demonstrates, however, the
qualitatively different behaviors of these models.

In the Anderson model, the ground state is composed of
charge complexes only. The toy model with the power $r=2$
exhibits a critical value $\delta_{cr}$ of the parameter
$\delta$ describing an unconventional behavior of the
density of states given in Eq. (1). At $\delta>\delta_{cr}$
the spectrum of charge complexes contains a gap.
If the gap is quite small, $\Delta<|Q|$, it does not effect
on the ground state properties, however its appearance changes
the thermodynamics of the system. A large gap, $\Delta>|Q|$,
not only changes the thermodynamic properties, but reconstructs
also the ground state of the system, because a part of charge
complexes decay into unbounded charge and spin excitations.
The behavior of the impurity level population is drastically
changed, however the Kondo effect takes place at any gap.

Finally, it should be emphasized that the toy versions are
interesting not only as exactly solvable examples of
unconventional magnetic alloys. It can be shown \cite{R3}
that an exact BA solution of a model with the density of
states $\rho(\epsilon)$ given in Eq. (1) and an energy
independent hybridization, $t=\mbox{const}$, exhibits
many of qualitative features described above.

I thank S. John, V. Yudson, and M. Zhitomirsky for stimulating
discussions.


\begin{references}

\bibitem[*]{em}
Electronic address: rupasov@physics.utoronto.ca
\vspace{3ex}

\bibitem{WF}
D. Withoff and E. Fradkin, \prl {\bf 64}, 1835 (1990).

\bibitem{BH}
L. S. Borkowski and P. J. Hirschfeld, \prb {\bf 46},
9274 (1992).

\bibitem{CF}
C. R. Cassanello and E. Fradkin, \prb {\bf 53}, 15079 (1996);

\bibitem{CJ}
K. Chen and C. Jayaprakash, J. Phys.: Condens. Matter,
{\bf 7}, L491 (1995).

\bibitem{I}
K. Ingersent, \prb {\bf 54}, 11936 (1996).

\bibitem{GBI}
C. Gonzalez-Buxton and K. Ingersent, \prb {\bf 54}, 15614 (1996);
preprint cond-mat/9803256.

\bibitem{BPH}
R. Bulla, Th. Pruschke, and A. C. Hewson, J. Phys.: Condens. Matter
{\bf 9}, 10463 (1997).

\bibitem{S}
T. Saso, J. Phys. Soc. Jpn. {\bf 61}, 3439 (1992).

\bibitem{SO}
T. Saso and J. Ogura, Physica B {\bf 186-188}, 372 (1993);

\bibitem{OS}
J. Ogura and T. Saso, J. Phys. Soc. Jpn. {\bf 62}, 4364 (1993).

\bibitem{TSS}
K. Takegahara, Y, Shimizu, and O. Sakai, J. Phys. Soc. Jpn.
{\bf 61} 3443 (1993).

\bibitem{TSGS}
K. Takegahara, Y, Shimizu, N. Goto, and O. Sakai, Physica B
{\bf 186-188}, 381 (1993).

\bibitem{CPC}
L. Cruz, P. Phillips, and A. H. Castro Neto, Europhys. Lett. {\bf 29},
389 (1995).

\bibitem{CJ2}
K. Chen and C. Jayaprakash, \prb {\bf 57}, 5225 (1998).

\bibitem{TW}
A. M. Tsvelick and P. B. Wiegmann, Adv. Phys. {\bf 32}, 453 (1983).

\bibitem{AFL}
N. Andrei, K. Furuya, and J. H. Lowenstein, \rmp {\bf 55}, 331 (1983).

\bibitem{PS}
P. Schlottmann, Phys. Rep. {\bf 181}, 1 (1989).

\bibitem{AH}
A. C. Hewson, {\em The Kondo Effect to Heavy Fermions}
(Cambridge University Press, Cambridge, 1993).

\bibitem{W}
K. G. Wilson, \rmp {\bf 47}, 773 (1975).

\bibitem{KM}
H. R. Krishna-Murthy, J. W. Wilkins, and K. G. Wilson,
\prl {\bf 35}, 1101 (1975); \prb {\bf 21}, 1003 (1980);
{\em ibid.}, {\bf 21} 1044 (1980).

\bibitem{A}
A. A. Abrikosov, Physica {\bf 2}, 5 (1965).

\bibitem{H}
F. D. M. Haldane, \prl {\bf 40}, 416 (1978).

\bibitem{R}
V. I. Rupasov, Phys. Lett. A {\bf 237}, 80 (1997).

\bibitem{RS}
V. I. Rupasov and M. Singh, J. Phys. A {\bf 29}, L205 (1996);
\prl {\bf 77}, 338 (1996); \pra {\bf 54}, 3614 (1996).

\bibitem{R2}
V. I. Rupasov, \prl {\bf 80}, 3368 (1998).

\bibitem{PW}
P. B. Wiegmann, Phys. Lett. A {\bf 80}, 163 (1980).

\bibitem{PS2}
P. Schlottmann, Z. Phys. B {\bf 49}, 109 (1982), {\em ibid.}
{\bf 52}, 127 (1983).

\bibitem{R3}
V. I. Rupasov, {\em to be published}.

\end{references}
\end{document}